\documentclass[epj]{svjour}

\usepackage{graphics}
\usepackage{rotate}
\usepackage{epsfig}

\begin{document}

\title{Parallel dynamics of the asymmetric extremely diluted Ashkin-Teller 
		neural network}
\author{D. Boll\'e \and G. Jongen}
\institute{ Instituut voor Theoretische Fysica, K.U. Leuven, 
	B-3001 Leuven, Belgium
	\\ \email{desire.bolle@fys.kuleuven.ac.be}}

\date{}

\abstract{
The parallel dynamics of the asymmetric extremely diluted Ashkin-Teller neural
network is studied using signal-to-noise analysis techniques. 
Evolution equations for the order parameters are derived, both at zero and 
finite temperature. The retrieval properties of the network are discussed in 
terms of the four-spin coupling strength and the temperature. It is shown 
that the presence of a four-spin coupling enhances the retrieval quality.
\PACS{{64.60.Cn}{Order-disorder transformations; statistical mechanics of 
			model systems} \and 
	{75.10.Hk}{Classical spin models} \and
	{87.10.+e}{General, theoretical, and mathematical biophysics} \and
	{02.50.-r}{Probability theory, stochastic processes and statistics} 
     } }  

\maketitle

\section{Introduction}
\label{intro}

Recently, the equilibrium properties of the Ashkin-Teller neural network ({\sc
atnn}) have been studied in \cite{BK98,BK99}. 
The neurons of the {\sc atnn} are described by {\it two} Ising spins 
of different types. This allows the network to
store and to retrieve pairs of patterns. 
Therefore, more complicated information can be stored in the {\sc atnn}
than in the Hopfield model \cite{Ho82,AGS87}, 
e.g., the fore- and background of a picture. 
Every spin is connected to spins of the same type. In addition, 
the neurons are connected to each other. The connections linking the neurons 
are four-spin couplings, since they connect two pairs of spins, 
one pair per neuron. 
This allows the network to retrieve both patterns of a pair simultaneously. 
One can think of the model as a combination of two Hopfield models, 
each retrieving one of the patterns. The four-spin coupling is then a
connection between both models. 
The underlying idea is that the simultaneous retrieval of a pair of patterns
is easier than the independent retrieval of the patterns in the pair. 

There are various reasons for studying this model. 
The Ashkin-Teller spin glass is related to disordered systems where the
disorder evolves on a time scale that can be tuned \cite{Ca94}. 
The introduction of a neuron containing different types of spins 
is also neurobiologicaly
motivated by the fact that areas in the brain exist which react to two
different kinds of dependent stimuli in such a way that the response to
particular combinations of these stimuli is stronger than the response
to others \cite{KI93}. Finally, in neuropsychological studies on amnesia, 
it has become appreciated that
memory is composed of multiple separate systems which can store
different types of informations, e.g., information based on skills and  
informations based on specific facts or data \cite{Sq93}.

In \cite{BK98,BK99}, the thermodynamic and retrieval properties of the  
{\sc atnn} have been studied using replica-symmetric mean-field theory.
In the present paper, we analyse the parallel dynamics of the asymmetric
extremely diluted version of the model. 
Both the way how the system evolves to its equilibrium configuration and the
properties of the equilibrium configuration itself are subjects of interest. 
It is known \cite{BKS90,BJS99b} that the dynamics in symmetric architectures, 
even in the diluted case, is complicated 
in a non-trivial way because of correlations between the neuron states. These
correlations are caused by feedback loops and common ancestors. 
In contrast to the Hopfield model, where the dynamics has been solved taking 
into account all the correlations \cite{BJS98,BJS99a}, the presence
of two types of spins makes the analysis of the correlations 
in the {\sc atnn} very complicated. 
The underlying reason is the existence of two sources of correlations.
First, feedback loops appear due to the two-spin interaction,
as in the Hopfield model. Second, 
the four-spin coupling causes correlations between
spins of different type. 
Therefore, in order to arrive at a first insight in the
dynamics of the model, we limit ourselves to its asymmetric extremely diluted
version where all correlations
between the neuron states are eliminated \cite{DGZ87,BSVZ94}. 

Using standard signal-to-noise analysis techniques (see, e.g, refs. 
\cite{BJS99b,DGZ87,PZ91}), 
we find that the local field of the asymmetrically diluted {\sc atnn}
contains only a normally distributed part, besides the signal. 
As observed already for the asymmetric diluted Hopfield model
\cite{DGZ87}, the structure of the local field does not change
in time. This allows us to write down immediately 
the complete time evolution of the main overlaps. 

The rest of the paper is organised as follows. 
In the second section, we define the model as
an extension of the Hopfield model. We 
introduce parallel dynamics at arbitrary temperature, and define
the main overlaps (one for each type of spins)
as macroscopic measures for the retrieval quality. In
Section~\ref{section:dynamicsatnn}, 
we use signal-to-noise analysis techniques in order to
write down the evolution equations at arbitrary time. 
From the evolution equations, 
the fixed-point equations are obtained. 
These equations lead to the dynamical capacity-temperature diagram
presented in Section~\ref{secatnn:results}. Finally, we give some
concluding remarks in Section~\ref{secatnn:concl}.

\section{The model}
\label{section:modelATNN}
The {\sc atnn} 
is defined as a neural network consisting of $N$ neurons. 
Each of the neuron states is described by two spins 
with value $\sigma_i$ and
$s_i$ ($i=1,\dots,N$), both taken from the discrete set $\{-1,+1\}$. 
For each type, the spins $i$ and $j$ are coupled by a two-spin interaction
$J_{ij}^{(1)}$ 
and $J_{ij}^{(2)}$ respectively, while the neurons $i$ and $j$ are
coupled by a four-spin interaction $J_{ij}^{(3)}$. 
We assume no diagonal terms viz.\ $J_{ii}^{(y)}=0,~y=1,2,3$. 

A configuration of an {\sc atnn} consists out of a 
$\mbox{\boldmath $\sigma$}$- and ${\vec s}$-part viz.\
\begin{equation}
  \left(\mbox{\boldmath $\sigma$}(t)=\{\sigma_j(t)\},
       {\vec s}(t)    =\{s_j(t)\}\right);~j=1,\dots,N \,.
  \nonumber 
\end{equation}
Given such a configuration, we define three types of local fields: 
two Hopfield-like local fields which measure the incoming
signal to the spins $\sigma_i$ and $s_i$, caused by the spins of the same
type
\begin{eqnarray}
   h_{N,i}^{(1)}({\mbox{\boldmath $\sigma$}}(t))=
                \sum_{j=1}^N J_{ij}^{(1)}\sigma_j(t)
   \nonumber \\
   h_{N,i}^{(2)}({{\vec s}}(t))=
                \sum_{j=1}^N J_{ij}^{(2)}s_j(t)\,, 
  \label{eq:hatnn1} 
\end{eqnarray}
and, in addition, 
a local field which measures the incoming signal to neuron $i$,
caused by both spins of the other neurons
\begin{eqnarray}
  h_{N,i}^{(3)}({\mbox{\boldmath $\sigma$}}(t),{{\vec s}}(t))&=&
     \sum_{j=1}^N J_{ij}^{(3)}
			\sigma_j(t)s_j(t)\,.
\label{eq:hatnn}
\end{eqnarray}
In the sequel, we write the shorthand notation 
$h^{(x)}_{N,i}(t) \equiv 
h_{N,i}^{(x)}({\vec S}_x(t)),~x=1,2$ with 
${\vec S}_1(t)={\mbox {\boldmath $\sigma$}}(t)$, $\vec S_2(t)=\vec s(t)$, 
and 
$h^{(3)}_{N,i}(t) \equiv h_{N,i}^{(3)}({\mbox{\boldmath $\sigma$}}(t),
	{{\vec s}}(t))$.
The configuration $(\mbox{\boldmath $\sigma$}(0),{\vec s}(0))$ is chosen as 
input.
At temperature $T=1/\beta$, all neurons are updated in parallel
according to the transition probability 
\begin{eqnarray}
  &&\hspace{-.5cm}
    \Pr(\sigma_i(t+1)=\sigma|\mbox{\boldmath $\sigma$}(t),{\vec s}(t))
    	\nonumber\\
    &&=\frac12\left[ 1+\tanh\beta\sigma
	  	\left(h_{N,i}^{(1)}(t)+s_i(t)h_{N,i}^{(3)}(t)
		\right)
	    \right]
    \nonumber \\
    &&\hspace{-.5cm}
    \Pr(s_i(t+1)=s|\mbox{\boldmath $\sigma$}(t),{\vec s}(t))
  	\nonumber \\
    &&=\frac12\left[ 1+\tanh\beta s
	  	\left(h_{N,i}^{(2)}(t)+\sigma_i(t)h_{N,i}^{(3)}(t)
		\right) \right]\,.
\label{eq:upatnnstoch}
\end{eqnarray}
We assume hereby that both types of spins exhibit the same degree of
stochasticity. 
At zero temperature, this dynamics becomes deterministic and is given by 
\begin{eqnarray}
  S_{x,i}(t+1)&=&\mbox{\rm sign}
	  	\left(h_{N,i}^{(x)}(t)+S_{\tilde x,i}(t)h_{N,i}^{(3)}(t)
		\right)	
\label{eq:upatnndet}
\end{eqnarray}
with $x,\tilde x=1,2$ and $x\neq\tilde x$.
The $\mbox{\boldmath $\sigma$}$-spins
receive at each time input from the ${\vec s}$-spins and vice versa 
due to the term containing $h_{N,i}^{(3)}(t)$.

The aim of the network is to store simultaneously $p_1$ patterns
$\{\mbox{\boldmath $\xi$}^\mu\},~\mu=1,\dots,p_1$ 
in the $\mbox{\boldmath $\sigma$}$-part of the network and 
$p_2$ patterns $\{\mbox{\boldmath $\eta$}^\mu\},~\mu=1,\dots,p_2$ in the
${\vec s}$-part. All components of the patterns $\xi_i^\mu$  and
$\eta_i^\mu$  are i.i.d.r.v.\ taken from $\{-1,+1\}$ with zero mean 
$\left<\xi_i^\mu\right>=0=\left<\eta_i^\mu\right>$ 
and independent type by type 
$\left<\xi_i^\mu\eta_j^\nu\right>=0$ ($i,j=1,\dots,N$).
In order to store these embedded patterns, the two-spin couplings are chosen
according to the Hebb rule 
\begin{eqnarray}
  J_{ij}^{(1)}=\frac{J_1}{N} \sum_{\mu=1}^{p_1} \xi_i^\mu \xi_j^\mu	
  \label{eq:J1} \qquad 
  J_{ij}^{(2)}=\frac{J_2}{N} \sum_{\mu=1}^{p_2} \eta_i^\mu \eta_j^\mu	
  	\,.
  \label{eq:J2}
\end{eqnarray}
The four-spin interaction is, also in analogy to the Hebb-rule, defined as
\cite{BK98}
\begin{eqnarray}
  J_{ij}^{(3)}&=&\frac{J_3}{N} \sum_{\mu=1}^{p_3}
  	\gamma_i^\mu \gamma_j^\mu  \,.
\label{eq:Jatnn}
\end{eqnarray} 
Under the assumption of independent embedded patterns 
$\mbox{\boldmath $\xi$}^\mu$
and $\mbox{\boldmath $\eta$}^\mu$, we consider the following form
\begin{equation}
  \gamma_i^\mu=\xi_i^\mu \eta_i^\mu	\,.
\nonumber
\end{equation}
The patterns $\{\mbox{\boldmath $\gamma$}^\mu\},~\mu=1,\dots,p_3$ 
are then a set of i.i.d.r.v.\ taken from $\{-1,+1\}$ with
zero mean. In the literature, this choice of patterns is called the linked
case \cite{BK98}.

The variables $J_y,~y=1,2,3$ are constants scaling the relative
importance of all types of couplings. 
We choose in the sequel $J_1=J_2$ since 
we want both types of spins interchangeable for simplicity.
The relative scale of
the temperature and coupling strengths is fixed by choosing $J_1=1$. 
Finally, we define $J=J_3/J_1$ such that this quantity measures the 
relative strength of the
four-spin couplings with respect to the two-spin couplings. 
In the limit $J\to 0$ the {\sc atnn} becomes, at least in structure, the
equivalent of two independent Hopfield models 
since $h_{N,i}^{(3)}(t)=0$ at all times. 
We use the temperature $T$ and the relative coupling
strength $J$ as independent variables. 

In the sequel, we take the interactions asymmetric extremely diluted 
\cite{DGZ87,BSVZ94}
\begin{eqnarray}
  \tilde J_{ij}^{(y)}=c_{ij}^{(y)}~N~J_{ij}^{(y)}/c_y
  \qquad y=1,2,3
\label{eq:modJdilatnn}
\end{eqnarray}
with $c_y>0$  and $\Pr\{c_{ij}^{(y)}=a\}=(1-c_y/N)\delta_{a,0} + 
(c_y/N) \delta_{a,1}$. 
The variables $\{c_{ij}^{(y)}\}$ are independent for
each pair $(i,j)$ representing both the asymmetry and the dilution. 
The diagonal terms are excluded $c_{ii}^{(y)}=0$.
The structure of the architecture of the network then becomes a directed
tree with an average number of incoming and outgoing connections 
(type by type) both equal to $c_y$.
It is assumed that $c_y\ll N$. 
The system is first diluted by taking the limit $N\to\infty$. Afterwards, 
the number of incoming signals per site is made
extensive by taking the limit $c_y\to\infty$. 
The probability to have feedback in the system is now zero and the 
correlations are treelike. 

In
principle, all couplings can be diluted independently (viz.\ $c_1\neq
c_2 \neq c_3$). For convenience, however, we dilute them in the same way,  
\begin{equation}
  c_{ij}^{(1)}=c_{ij}^{(2)}=c_{ij}^{(3)}\equiv c_{ij} 
  \qquad c_1=c_2=c_3\equiv c \,.
   \nonumber
\end{equation}
This means that both spins of a neuron get information from the same neurons
and that the dynamics of both types of spins can be treated analogously.
Since the number of embedded patterns is of the same order as 
the number of connections a spin has with spins of the same type, we have 
$p_1=p_2=p_3=\alpha c\equiv p$.
In what follows, we write, for simplicity, $J_{ij}^{(y)}$ instead of
$\tilde{J}_{ij}^{(y)}$. 

At this point, we note that the capacity of the {\sc atnn} is defined as the
ratio of the number of patterns stored in the network and the number of
couplings to a neuron. In this model where we want to store $2p$ patterns
$\{\mbox{\boldmath $\xi$}^\mu,\mbox{\boldmath $\gamma$}^\mu\}$, 
all neurons have in average $3c$ links to the other
neurons: $2c$ two-spin couplings and $c$ four-spin couplings. 
Therefore, the capacity of the {\sc atnn} equals
\begin{equation}
  \alpha_{\rm ATNN}\equiv\frac{2p}{3c}=\frac23\alpha \,.
  \nonumber
\end{equation}

The retrieval quality of the model is measured by the Hamming distance
between the microscopic state of the network and the stored patterns
\begin{eqnarray}
  &&D^\mu_H(t)\equiv\left[\sum_{x}
  	\left(d_x({\mbox{\boldmath $\psi$}}_x^\mu,{\vec S}_x(t))\right)^2
	\right]^{\frac12}
  \label{eq:hamdistatnn1}
  \\
  &&d_x({\mbox{\boldmath $\psi$}}_x^\mu,{\vec S}_x(t))\equiv
                \frac{1}{N}
                \sum_{i=1}^N[\psi_{x,i}^\mu-S_{x,i}(t)]^2
  \label{eq:hamdistatnn}
\end{eqnarray}
where $\mu=1,\dots,p$, 
$x=1,2$, ${\mbox{\boldmath$\psi$}}_1^\mu={\mbox{\boldmath$\xi$}}^\mu$
and ${\mbox{\boldmath$\psi$}}_2^\mu={\mbox{\boldmath$\eta$}}^\mu$. 
This naturally introduces the main overlaps 
\begin{eqnarray}
        m_{x,N}^\mu(t)=\frac{1}{N}
                \sum_{i=1}^N\psi_{x,i}^\mu S_{x,i}(t)
                \quad \mu =1,\dots,p \,.
\label{eq:matnndef}
\end{eqnarray}
In the diluted model the sum in (\ref{eq:matnndef}) 
has to be taken over  the tree-like structure, viz.
$\frac1N\sum_{i=1}^N \to \frac1c\sum_{i=1}^Nc_{ij}$. The expression for
the  main overlap (\ref{eq:matnndef}) then reads
\begin{eqnarray}
        m_{x,c,N}^\mu(t)=\frac{1}{c}
                \sum_{i=1}^Nc_{ij}\psi_{x,i}^\mu S_{x,i}(t)
                \quad \mu =1,\dots,p \,.
\label{eq:matnndefc}
\end{eqnarray}
We remark that both expressions (\ref{eq:matnndef}) and (\ref{eq:matnndefc}) 
become equal in the thermodynamic limit $c,N\to\infty$.

\section{Dynamics}
\label{section:dynamicsatnn}

In this section we construct a set of recursion equations for the main
overlap  order parameters. We use hereby signal-to-noise techniques     
(see, e.g., refs. \cite{BSVZ94,PZ91}).
Finally, we write down the fixed-point equations. 

Suppose an initial spin configuration 
$(\mbox{\boldmath $\sigma$}(0),{\vec s}(0))$. The configurations 
${\vec S}_x(0)=\{S_{x,i}(0)\},~i=1,\dots,N$
are collections of i.i.d.r.v.\ with mean
$\left<S_{x,i}(0)\right>=0$ and variance 
$\left<\left(S_{x,i}(0)\right)^2\right>=1$. 
Spins of different types are uncorrelated 
$\left<\sigma_i(0)s_j(0)\right>=0$ ($i,j=1,\dots,N$). 
Both types are correlated with only one of the stored patterns, e.g., the
first one 
\begin{eqnarray}
     \left<\psi_{x,i}^\mu S_{x,j}(0)\right>= 
                         \delta_{i,j}\delta_{\mu,1}m^1_{x,0}
     	\,.
\label{eq:initatnnsigma}
\end{eqnarray}
The site by site independence of spins and patterns implies by the law
of large numbers ({\sc lln}) that we get for the main overlaps
\begin{eqnarray}
   m^1_x(0)\equiv\!\lim_{c,N\to\infty}m_{x,c,N}^1(0)
     =\left<\psi_{x,i}^1S_{x,i}(0)\right>=m^1_{x,0} \,.
\label{eq:matnn(0)}
\end{eqnarray}

We now want to study how the main overlaps evolve under the parallel
dynamics specified before. For a general time step and at $T=0$, we find
from (\ref{eq:matnndefc}) and the {\sc lln} in the limit $c,N\to\infty$  
\begin{eqnarray}
  \label{eq:matnn}
  m_x^1(t+1) &=&
    \left<\!\left<
      \psi_{x,i}^1 \mbox{\rm sign} 
      	\left(h_i^{(x)}(t)+\psi_{\tilde x,i}(t) h_i^{(3)}(t)\right)
	\right>\!\right> \nonumber\\     
  m_3^1(t+1) &=&
    \left<\!\left<
      \xi_i^1 \eta_i^1
      \mbox{\rm sign} \left(h_i^{(1)}(t)+s_i(t) h_i^{(3)}(t)\right)
      	\right.\right.\nonumber \\ &&\qquad\left.\left.
      \times\mbox{\rm sign} \left(h_i^{(2)}(t)+
                           \sigma_i(t) h_i^{(3)}(t)\right)
	\right>\!\right> 
\end{eqnarray}
where $x,\tilde x=1,2;~x\neq\tilde x $. 
The average  $\left<\!\left< \cdot \right>\!\right>$
denotes the average both over the distribution of the embedded patterns
$\{\xi_i^\mu\}$ and $\{\eta_i^\mu\}$ and the initial configuration
$\{\sigma_i(0),s_i(0)\}$. The average over the latter is 
hidden in an average over the local fields through the updating rule 
(\ref{eq:upatnnstoch}). 

The equations (\ref{eq:matnn}) show that the knowledge of the
distribution of the local field at successive time steps is sufficient
in order to find the evolution equations for the order parameters. 
We start with calculating the distribution of
the local field of the $\mbox{\boldmath $\sigma$}$-spins at $t=0$. 
Using the definitions (\ref{eq:Jatnn}) and applying 
the signal-to-noise analysis, we have 
\begin{equation}
  h_{N,i}^{(1)}(0)=\xi_i^1\frac 1c \sum_{j=1}^N c_{ij}
	\xi_j^1 \sigma_j(0)
    + \frac 1c \sum_{\mu=2}^p \xi_i^\mu
	\sum_{j=1}^Nc_{ij} \xi_j^\mu \sigma_j(0)	\,.
  \label{eq:h0a}
\end{equation}
The signal term, i.e., the first term on the r.h.s.\ of (\ref{eq:h0a}),
is nothing but
the main overlap (\ref{eq:matnndefc}) multiplied by $\xi_i^1$. 
In the noise part, i.e., the second term on the r.h.s., 
all terms are uncorrelated by construction such that we can apply the 
central limit theorem ({\sc clt}) to find 
\begin{equation}
  \lim_{c,N\to\infty} \frac{1}{\sqrt{p}} 
    \sum_{\mu=2}^p \xi_i^\mu \frac{1}{\sqrt{c}}
    \sum_{j=1}^N c_{ij} \xi_j^\mu \sigma_j(0)
    \sim {\cal N}(0,1) 	
  \nonumber
\end{equation}
where ${\cal N}(0,1)$ represents a Gaussian random variable with mean
$0$ and variance $1$.
Therefore, in the limit $c,N\to\infty$, the local field at $t=0$ is the
sum of  two independent random variables 
\begin{eqnarray}
  h_i^{(1)}(0)\equiv\lim_{c,N\to\infty}h_{N,i}^{(1)}(0)
  	=\xi_i^1 m_1^1(0) + \sqrt\alpha ~z_1(0)
  \label{eq:h100}
\end{eqnarray}
with $z_1(0)\sim{\cal N}(0,1)$. In an analogous way, we find for the local 
field of the ${\vec s}$-spins 
\begin{eqnarray}
  h_i^{(2)}(0)\equiv\lim_{c,N\to\infty}h_{N,i}^{(2)}(0)
  	=\eta_i^1 m_2^1(0) + \sqrt\alpha ~z_2(0)
\label{eq:hatnnsna}
\end{eqnarray}
with $z_2(0)\sim{\cal N}(0,1)$. 

As in the local fields of the spins (\ref{eq:hatnn1}), 
we separate in the local field (\ref{eq:hatnn}) the
terms containing the first pattern 
$\mbox{\boldmath $\gamma$}^1$ from the rest
\begin{eqnarray}
  h_{N,i}^{(3)}(0)&=&\gamma_i^1\frac 1c \sum_{j=1}^N c_{ij}
	\gamma_j^1 \sigma_j(0) s_j(0)
	\nonumber \\
    &&\quad + \frac 1c \sum_{\mu=2}^p \gamma_i^\mu
	\sum_{j=1}^N c_{ij} \gamma_j^\mu \sigma_j(0)s_j(0)	\,.
\end{eqnarray}
In analogy with before, we call the first term the signal and the last
term the noise. Applying the {\sc lln} to the signal term, we get 
\begin{equation}
  \lim_{c,N\to\infty}
  \frac 1c \sum_{j=1}^N c_{ij} \gamma_j^1 \sigma_j(0) s_j(0)
    = \left<\gamma_j^1 \sigma_j(0) s_j(0)\right> \,.
  \nonumber
\end{equation}
Since this term resembles strongly the main overlaps
(\ref{eq:matnndefc}), we call it also an overlap and denote
it by $m_{3}^1(0)$. In general, this overlap is defined by 
\begin{equation}
        \label{eq:mdef3}
        m_{3,c,N}^\mu(t)=\frac{1}{c}
                \sum_{i=1}^N c_{ij}\gamma_i^\mu\sigma_i(t) s_i(t)
                \quad \mu =1,\dots,p \,.
\end{equation}
In the sequel, we will treat this parameter at
the same level as the other overlaps (\ref{eq:matnndefc}). 
For the linked choice of patterns
$\gamma_i^\mu\equiv\xi_i^\mu \eta_i^\mu$, it follows from 
(\ref{eq:initatnnsigma})
that $\left<\gamma_i^\mu\sigma_j(0)s_k(0)\right>=
\delta_{ij}\delta_{ik}\delta_{\mu1}
m^1_{1,0}m^1_{2,0}$ since the initial spin configurations 
$\sigma_j(0)$ and
$s_j(0)$ are independent. Therefore, we have in the thermodynamic limit 
and at $t=0$
\begin{equation}
  \label{eq:m3(0)}
  m_3^1(0)\equiv \lim_{c,N\to\infty}m_{3,c,N}^1(0)=m_1^1(0) m_2^1(0)\,.
\end{equation}
Following the same line of arguments as before, the noise term converges
again to a Gaussian random variable such that
\begin{equation}
  h_i^{(3)}(0)\equiv\lim_{c,N\to\infty}h_{N,i}^{(3)}(0)
  	= J\xi_i^1\eta_i^1 m_3^1(0) + J\sqrt\alpha ~ z_3(0)
  \label{eq:h300}
\end{equation}
with $z_3(0)\sim{\cal N}(0,1)$.
This finishes the calculation of the local fields at time $t=0$.

At a general time $t$, the local fields still consist out of a signal
term, proportional to the main overlap, and a Gaussian distributed noise
part.  This is due to the extreme dilution which eliminates all common
ancestors in the dynamics. Therefore, all variables
$\{X_j^\mu\equiv\psi_{x,i}^\mu c_{ij}\psi_{x,j}^\mu S_{x,j}(t)|
	j=1,\dots,N;~\mu=1,\dots,p\}$
are a set of i.i.d.r.v.\ and we can apply the {\sc clt} in the same way as in
eq.~(\ref{eq:h0a}). So, we find for the distribution of
the local fields a set of equations with the same structure as the 
eqs.~(\ref{eq:h100}), (\ref{eq:hatnnsna}) and (\ref{eq:h300}), viz.\
\begin{eqnarray}
  h_i^{(x)}(t) &=& \psi_{x,i}^1 m_x^1(t)+\sqrt\alpha~z_x(t)
     ;~ z_x(t)\sim{\cal N}(0,1)
  \label{eq:hatnn(t)}
  \nonumber \\
  h_i^{(3)}(t) &=& J\xi_i^1 \eta_i^1 m_3^1(t)+
  	J\sqrt\alpha~z_3(t)
	;~ z_3(t)\sim{\cal N}(0,1)\,.
    \nonumber \\
\end{eqnarray}
The three Gaussian variables $z_y(t)$ are uncorrelated. 

Using the distributions of the local fields (\ref{eq:hatnn(t)}) and 
remarking that the joint probability of $s_i(t),\sigma_i(t),\xi_i^1$ and 
$\eta_i^1$ is
obtained from the overlaps $m_y^1(t),~y=1,2,3$, 
the equations (\ref{eq:matnn}) and (\ref{eq:hatnn(t)})
lead immediately to the evolution equations for the order parameters at 
zero temperature 
\begin{eqnarray}
  m_x^1(t+1)&=&\sum_{\sigma=\pm1} 
    \frac 12 \left(1+\sigma m_{\tilde x}^1(t)\right)
    \mbox{\rm Erf}\left(\frac{m_x^1(t)+\sigma J m_3^1(t)}
    	{\sqrt{2\alpha(1+J^2)}}\right)
  \nonumber  \\
  m_3^1(t+1)&=&\sum_{\sigma,\tau=\pm1} 
    \left(\sigma \tau +\sigma m_1^1(t) + \tau m_2^1(t) + m_3^1(t)\right) 
    \nonumber \\ && \qquad
    \int {\cal D} z ~
    \mbox{\rm Erf}\left(\frac{\sigma m_1^1(t)+Jm_3^1(t)+J\sqrt{\alpha}z}
    		{\sqrt{2\alpha}}\right)
    \nonumber \\ &&\qquad
    \times\mbox{\rm Erf}\left(\frac{\tau m_2^1(t)
                 +J m_3^1(t)+J\sqrt{\alpha}z}{\sqrt{2\alpha}}\right)	
    \label{eq:matnn(t)final}
\end{eqnarray}
where $x,\tilde x=1,2,~ x\neq\tilde x $.

After some time $t$ the dynamics reaches the point where the spins 
macroscopically equilibrate. This means that the main overlap becomes
stationary, viz.\ $m^1_y(t+1)=m^1_y(t)$.
Since the expressions for the local fields do not
change their structure, the corresponding fixed-point equations are 
easily obtained from
eq.~(\ref{eq:matnn(t)final}) by replacing the time dependent quantities
by  their equilibrium value $m_y^1\equiv\lim_{t\to\infty}m_y^1(t)$. 
In the limit $J\to0$, the equations (\ref{eq:matnn(t)final}) are
consistent  with the evolution equations of \cite{DGZ87}.

At non-zero temperature $T=1/\beta$ the main overlaps at time~$t$ read 
\begin{eqnarray}
  m_x^1(t+1)&=&\left<\!\left<
      \psi_{x,i}^1 \left< S_{x,i}(t+1)\right>_\beta
    \right>\!\right>
  \nonumber \\
  m_3^1(t+1)&=&\left<\!\left<
      \xi_i^1 \eta_i^1 \left< \sigma_i(t+1)\right>_\beta
		       \left< s_i(t+1)\right>_\beta
    \right>\!\right> \,.
\label{eq:matnnstoch(t)}
\end{eqnarray}
The thermal averages are defined by the updating rule
(\ref{eq:upatnnstoch}) and can be written as 
\begin{eqnarray}
  \hspace{-.6cm}
  \left< S_{x,i}(t+1)\right>_\beta&=&
    \tanh\left[\beta\left(h_i^{(x)}(t)+
    	S_{\tilde x,i}(t)h_i^{(3)}(t)\right)\right]\,.
\label{eq:thermaverageatnn}
\end{eqnarray}
The stochasticity in the dynamics does not modify the local spin
correlations when compared with the deterministic dynamics.
Therefore, the local fields are still distributed according to
(\ref{eq:hatnn(t)}) and we get for the order parameters 
\begin{eqnarray}
  m_x^1(t+1)&=&\sum_{\sigma=\pm1} \frac12 
  	\left(1+\sigma m_{\tilde x}^1(t)\right)
    \nonumber \\ && \hspace{-1.8cm} \times
    \int \!{\cal D} y  \tanh \beta 
	\left( m_x^1(t) + \sigma J m_3^1(t) + \sqrt{\alpha(1+J^2)}y
	\right)
  \nonumber \\ 
  \nonumber \\ 
  m_3^1(t+1)&=&\sum_{\sigma,\tau=\pm1} 
    \frac14 \left(\sigma\tau +\sigma m_1^1(t)+ \tau m_2^1(t) 
              + m_3^1(t)\right)
                            \nonumber \\ 
    && \hspace{-1.8cm} \times
    \int \!{\cal D} z\!\int \!{\cal D} x  \tanh \beta 
	\left( \sigma m_1^1(t) + J m_3^1(t) + \sqrt{\alpha}x 
	+ J\sqrt{\alpha}z \right)
    \nonumber \\ && \hspace{-1.8cm} \times
 	      \int \!{\cal D} y  \tanh \beta 
	\left( \tau m_2^1(t) + J m_3^1(t) + \sqrt{\alpha}y 
	+ J\sqrt{\alpha}z \right)\,.
\label{eq:matnn(t)stoch}
\end{eqnarray}
The fixed-point equations are read off from (\ref{eq:matnn(t)stoch}) by
using again $m_y^1=\lim_{t\to\infty}m_y^1(t)$, $y=1,2,3$. In
the zero-temperature limit $\beta\to\infty$, the equations above
reduce to (\ref{eq:matnn(t)final}).
Moreover, in the limit $J\to0$, they are consistent with the ones 
obtained for the asymmetric 
extremely diluted Hopfield model in \cite{DGZ87}.

\section{Results}
\label{secatnn:results}
In this section, we discuss the numerical results for the {\sc atnn} 
obtained from the
fixed-point equations derived in the previous section. We present the
capacity-temperature diagram indicating the regions of retrieval as a 
function of the capacity $\alpha_{\rm ATNN}$ and the temperature $T$, and 
some representative figures illustrating the main features of
the model.

\begin{figure}
\epsfxsize=7.5cm
\centerline{\hspace{-.5cm}
	\rotate[r]{\epsfbox{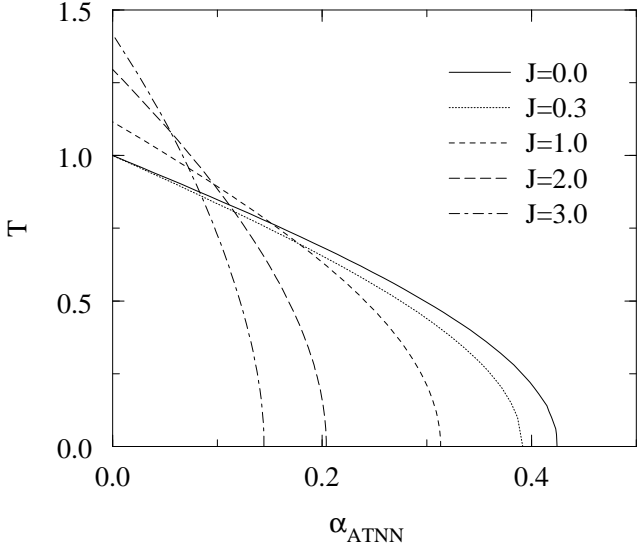}}}
\caption{Capacity-temperature diagram for the {\sc atnn} for J=0.0 (full
line),  $J=0.3$ (dotted line), $J=1.0$ (dashed line), $J=2.0$ (long
dashed  line) and $J=3.0$ (dot-dashed line).}
\label{figatnn:phasediag}
\end{figure}

Due to the choice $J_1=J_2=1$ and due to the condition that both types
of spins  have a finite initial overlap with one condensed pattern, it
turns  out that the overlaps $m_1(t)$ and $m_2(t)$ always converge to
the same  equilibrium values, independent of the size of the initial
overlaps.  (We forget about the superscript $1$). Therefore we can
restrict  ourselves to the case $m_1=m_2$. 

The resulting capacity-temperature diagram is pre\-sen\-ted in
Fig.~\ref{figatnn:phasediag}. 
First, we consider the special case $J=0$. At $T=0$
a non-zero solution for the fixed-point equations exists as long as 
$\alpha<2/\pi$,
indicating a transition from the retrieval to the non-retrieval regime 
at $\alpha_{\rm ATNN}=4/3\pi$. When the temperature
increases, the critical capacity decreases to become zero at $T=1$. 
The resulting transition line in the capacity-temperature 
diagram is similar to the one of the Hopfield model
\cite{DGZ87} up to a rescaling of the capacity. This is not surprising  
since the structure of the equations of the {\sc atnn} 
for $J=0$ is consistent with those of the
Hopfield model. The transition is always continuous, and 
the main overlap decreases when $\alpha_{\rm ATNN}$ increases. 
This indicates 
that the more embedded patterns, the harder the retrieval and the worse
the retrieval quality. 

\begin{figure*}
\epsfxsize=6.cm
\centerline{\rotate[r]{\epsfbox{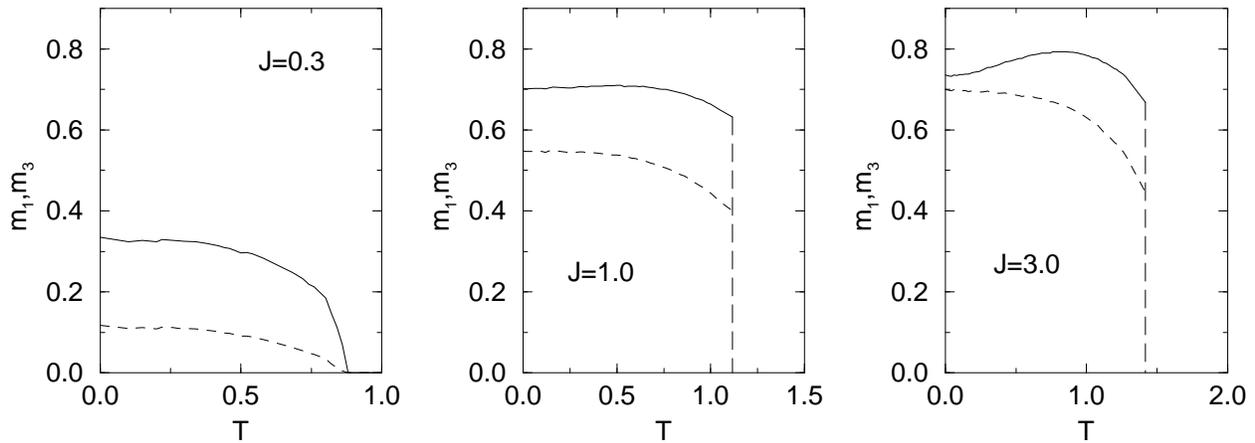}}}
\caption{Overlaps $m_1$ (full line) and $m_3$ (broken line) at the critical
capacity as a function of the temperature, 
for $a$) J=0.3, $b$) J=1.0 and $c$) J=3.0.}
\label{figatnn:overlap}
\end{figure*}

When the four-spin coupling is non-zero, things become different. A larger 
four-spin coupling makes retrieval possible at higher temperature and
decreases  the critical capacity at low temperature. 

For finite loading ($\alpha_{\rm ATNN}=0$), a continuous transition occurs 
for $J\leq1/3$ at $T=1$. For larger $J$, the transition becomes
discontinuous  and the critical temperature becomes larger. This
indicates that  a model with non-zero  four-spin couplings can perform
better  in the presence of noise in the dynamics. 
This corresponds with the results in \cite{BK98}. The overlap $m_1$
at the critical temperature
increases up to $J=2$, meaning that the larger $J$ the better the retrieval
quality. From $J=2$ onwards, the increasing noise in the 
dynamics at the transition line results in a slowly decreasing
overlap $m_1$. We note that we always observe $m_3=(m_1)^2$ when
$\alpha_{{\rm ATNN}}=0$. 

At zero temperature ($T=0$), 
increasing $J$ implies decreasing the critical capacity. The
transition is always first order, except for $J=0$. The
main overlap $m_1$ at the transition line first increases with $J$, but
starts to decrease from $J=2.0$ onwards. At larger $J$, it starts
increasing  again. The overlap $m_3$, however, 
always increases and becomes larger than $m_1$ for $J\geq4.2$.
The critical capacity for $J=1$ is equal to $0.3131$, which is higher
than  that of the fully connected {\sc atnn} ($\alpha_c=0.1839$)
\cite{BK99}.  This is consistent with the results obtained by comparing
the  asymmetric extremely
diluted with the fully connected Hopfield model \cite{AGS87,DGZ87}.

For non-zero temperature and infinite loading, the transition is 
partially continuous as long as $J\in[0,1/3]$. 
The larger $J$, the larger the temperature is where a 
continuous transition occurs. As an example, we have drawn the value of the 
overlaps at the critical capacity for $J=0.3$
(Fig.~\ref{figatnn:overlap}$a$) where the transition is continuous for 
$T\geq0.88$. 
When $J\geq1/3$, the transition is discontinuous for all temperature. In 
Fig.~\ref{figatnn:overlap}$b$ and Fig.~\ref{figatnn:overlap}$c$,
we have drawn the overlaps at the critical capacity for $J=1$ and $J=3$. 
The overlap $m_1$ exhibits a maximum at $T=0.32$ and $T=0.86$ respectively 
while $m_3$ is always decreasing.

\section{Concluding remarks}
\label{secatnn:concl}

In this article, we have studied 
the parallel dynamics of the asymmetric extremely diluted {\sc atnn} 
with linked patterns at arbitrary temperature. 
Because of the absence of correlations between the neurons, we have found 
that the noise of the local field at all time steps is normally
distributed.  
Hence, the dynamical equations for the order parameters are obtained 
immediately. Furthermore, the dynamical capacity-temperature diagram is 
discussed. 

In the presence of the four-coupling term, the dynamics can 
exhibit more noise without disturbing the retrieval process completely. 
Moreover, the transition from the retrieval to the non-retrieval regime  
becomes first order. This implies that the Hamming
distance becomes smaller, even at the transition line. 
So in general, we can say that the four-coupling term enhances the
retrieval  quality of the network. 

\begin{acknowledgement}
The authors are indebted to L.~Bianchi for some contributions at the
initial stages of this work and to P.~Koz\l owski for constructive
discussions. One of us (D.B.) thanks the Fund for Scientific 
Research - Flanders  (Belgium) for financial support.
\end{acknowledgement}


\begin{thebibliography}{}
\bibitem{BK98} D.~Boll\'e and P.~Koz{\l}owski, 
   J.~Phys.~A: Math. Gen. {\bf 31}, 6319 (1998).
\bibitem{BK99} D.~Boll\'e and P.~Koz\l owski,
   {\tt cond-mat/9906274}, (to be published in J.~Phys.~A: Math. Gen.)
   (1999). 
\bibitem{Ho82} J.J.~Hopfield, 
   Proc.~Nat.~Acad.~Sci.~USA {\bf 79}, 2554 (1982).
\bibitem{AGS87} D.~Amit, H.~Gutfreund and H.~Sompolinsky, 
   Ann.~Phys. (N.~Y.) {\bf 173}, 30 (1987).
\bibitem{Ca94} N.~Caticha, 
   J.~Phys.~A: Math. Gen. {\bf 27}, 5501 (1994).
\bibitem{KI93} H. Komatsu and Y. Ideura,
   J.~Neurophysiol. {\bf 70}, 677 (1993).
\bibitem{Sq93} L.R. Squire, 
   {\it Brain mechanisms of perception and memory}
   ed. T.~Ono {\it et al.} 
   (Oxford University Press, 1993), 219;
   Science {\bf 232}, 1612 (1986).
\bibitem{BKS90} E.~Barkai, I.~Kanter and H.~Sompolinsky,
   Phys. Rev. A {\bf 41}, 590 (1990).
\bibitem{BJS99b} D.~Boll\'e, G.~Jongen and  G.M.~Shim,
   to appear in the {\it Proceedings of the International Conference on 
   Mathematical Physics and Stochastic Analysis (Lisbon, October 1998)},
   ed.  S.~Albeverio {\it et al.} (World Scientific, 1999). 
\bibitem{BJS98} D.~Boll\'e, G.~Jongen and G.M.~Shim,
   J.~Stat.~Phys. {\bf 91}, 125 (1998).
\bibitem{BJS99a} D.~Boll\'e, G.~Jongen and  G.M.~Shim,
   J.~Stat.~Phys. {\bf 96}, 861 (1999). 
\bibitem{DGZ87} B.~Derrida, E.~Gardner, and A.~Zippelius,
   Europhys.~Lett. {\bf 4}, 167 (1987).
\bibitem{BSVZ94} D.~Boll\'e, G.M.~Shim, B.~Vinck, and V.A.~Zagrebnov,
   J.~Stat.~Phys. {\bf 74}, 565 (1994).
\bibitem{PZ91} A.E.~Patrick and V.A.~Zagrebnov,
   J.~Phys.~A: Math. Gen. {\bf 24}, 3413 (1991); 
   J.~Stat.~Phys. {\bf 63}, 59 (1991).
\end{thebibliography}
\end{document}